\documentclass[longauth]{aa}  
\usepackage{graphicx}
\usepackage{color}
\usepackage[varg]{txfonts}
\usepackage{natbib}
\usepackage{amssymb}
\pdfoutput=1
\begin{document}

   \title{ALMA imaging of C$_2$H emission in the disk of \object{NGC\,1068}}

 \author{S.~Garc\'{\i}a-Burillo \inst{1}
			\and
	   S.~Viti\inst{2}		
			\and		
	   F.~Combes\inst{3}		
			\and	
	   A.~Fuente\inst{1}
	  		\and				
	   A.~Usero\inst{1} 		   
			\and
	  L.~K.~Hunt\inst{4} 
			\and		
	  S.~Mart\'{\i}n\inst{5,6} 		   
			\and	
	  M.~Krips	\inst{7} 		   
			\and			
	  S.~Aalto\inst{8}
	  		\and
	  R.~Aladro\inst{8, 9}		
   		\and	
	  C.~Ramos Almeida\inst{10, 11}
	  	\and	
	  A.~Alonso-Herrero\inst{12}		
	        \and      
	  V.~Casasola\inst{4}   
  	 	 \and       		
	  C.~Henkel\inst{9, 13}
	 	\and
      	  M.~Querejeta\inst{1,14}
	  	\and
	  R.~Neri\inst{7}
	 	\and	
	  F.~Costagliola\inst{8}	 
		\and
	  L.~J.~Tacconi\inst{15}
	        \and
	 P.~P.~van der Werf\inst{16}	  
	  }
   \institute{
          Observatorio Astron\'omico Nacional (OAN-IGN)-Observatorio de Madrid, Alfonso XII, 3, 28014-Madrid, Spain 
			  \email{s.gburillo@oan.es} 		  
      \and    
    Department of Physics and Astronomy, UCL, Gower Place, London WC1E 6BT, UK
     \and 
      Observatoire de Paris, LERMA, CNRS, 61 Av. de l'Observatoire, 75014-Paris, France 
	\and
	 INAF-Osservatorio Astrofisico di Arcetri, Largo Enrico Fermi 5, 50125-Firenze, Italy 	 
	\and
	Joint ALMA Observatory, Alonso de C\'ordova, 3107, Vitacura, Santiago 763-0355, Chile
	\and	
	ESO, Alonso de C\'ordova, 3107, Vitacura, Santiago 763-0355, Chile	 
	 \and	 
	 Institut de Radio Astronomie Millim\'etrique (IRAM), 300 rue de la Piscine, Domaine Universitaire de Grenoble, 38406-St.Martin d'H\`eres, France 
	 \and	 
	 Department of Earth and Space Sciences, Chalmers University of Technology, Onsala Observatory, 439 94-Onsala, Sweden  	 
	 \and
	 Max-Planck-Institut f\"ur Radioastronomie, Auf dem H\"ugel 69, 53121, Bonn, Germany	
	 \and	 
	Instituto de Astrof\'{\i}sica de Canarias, Calle V\'{\i}a L\'actea, s/n, E-38205 La Laguna, Tenerife, Spain
	\and
	Departamento de Astrof\'{\i}sica, Universidad de La Laguna, E-38205, La Laguna, Tenerife, Spain
	 \and
	Centro de Astrobiolog\'{\i}a (CSIC-INTA), ESAC Campus, 28692 Villanueva de la Ca\~nada, Madrid, Spain           
	 \and
	 Astronomy Department, King Abdulazizi University, P.~O. Box 80203, Jeddah 21589, Saudi Arabia
	\and
	European Southern Observatory (ESO), Karl-Schwarzschild-Strasse 2, D-85748 Garching bei M\"unchen, Germany
	\and	 
         Max-Planck-Institut f\"ur extraterrestrische Physik, Postfach 1312, 85741-Garching, Germany 
	\and
	Leiden Observatory, Leiden University, PO Box 9513, 2300 RA Leiden, Netherlands 
	}

   \date{Received--, 2017; ---, ---}

 
  \abstract
  {}
   {We study the feedback of star formation and nuclear activity on the chemistry of molecular gas in NGC~1068, a nearby ($D=14$~Mpc) Seyfert 2 barred galaxy, by analyzing if the abundances of key molecular species like ethynyl (C$_2$H), a classical tracer of Photon Dominated Regions (PDR), change in the different environments of the disk of the galaxy.}
   {We have used the Atacama Large Millimeter Array (ALMA) to map the emission of the hyperfine multiplet of C$_2$H($N=1-0$) and its underlying continuum emission in the central $r\simeq35\arcsec$(2.5~kpc)-region of the disk of NGC~1068  with a spatial resolution $1\farcs0\times0\farcs7$~($\simeq50-70$~pc). Maps of the dust continuum emission obtained at 349~GHz by ALMA have been used to derive the H$_2$ gas column densities and have been combined  with the C$_2$H map at matched spatial resolution to estimate the fractional abundance of this species. We have developed a set of time-dependent chemical models, which include shocks, gas-phase PDRs, and gas-grain chemical models, to determine the origin of the C$_2$H gas.}
   {A  sizeable fraction  of the total  C$_2$H line emission is detected from the $r\simeq1.3$~kpc starburst (SB) ring, a region that concentrates the bulk of the recent massive star formation in the disk  traced by the Pa$\alpha$ emission complexes imaged by the Hubble Space Telescope (HST). However, the brightest C$_2$H emission originates from a $r\simeq200$~pc off-centered circumnuclear disk (CND), where evidence of a molecular outflow has been previously found in other molecular tracers imaged by ALMA. We also detect significant emission that connects the CND with the outer disk in a region that probes the interface between the molecular disk and the ionized gas outflow out to $r\simeq400$~pc. 
We derived the fractional abundances of C$_2$H  ($X$(C$_2$H)) assuming  local thermodynamic equilibrium (LTE) conditions and a set of excitation temperatures ($T_{\rm ex}$) constrained by the previous multiline CO studies of the galaxy.  Our estimates range from $X$(C$_2$H)~$\simeq$~a few 10$^{-8}$  in the SB ring up to $X$(C$_2$H)~$\simeq$~ a few 10$^{-7}$ in the outflow region. PDR models that incorporate gas-grain chemistry are able to account for 
$X$(C$_2$H) in the SB ring for moderately dense ($n$(H$_2$)~$\geq$~10$^{4}$cm$^{-3}$) and moderately UV-irradiated gas (UV-field~$\leq$~10~$\times$~Draine field, where 1 Draine field~$\equiv$~2.74~$\times$~10$^{-3}$~erg~s$^{-1}$~cm$^{-2}$) in a steady-state regime, which depending on the initial and physical conditions of the gas may be achieved by 10$^5$~yr or as late as $10^7$~yr. However, the high fractional abundances estimated for C$_2$H in the outflow region can only be reached at {\it very early} times ($T\leq$ 10$^{2-3}$ yr) in models of UV/X-ray irradiated  dense gas ($n$(H$_2$)~$\geq$~10$^{4-5}$cm$^{-3}$).}
   {We interpret that the transient conditions required to fit the high values of $X$(C$_2$H) in the outflow are likely due to UV/X-ray irradiated non-dissociative shocks associated with the highly turbulent interface between the outflow and the molecular gas in NGC~1068.  Although the inferred {\it local} time-scales are short, the erosion of molecular clouds by the active galactic nucleus (AGN) wind and/or the jet likely resupplies continuously the interface working surface making a nearly steady state persist in the disk of the galaxy.}
    \keywords{Galaxies: individual: NGC\,1068 --
	     Galaxies: ISM --
	     Galaxies: kinematics and dynamics --
	     Galaxies: nuclei --
	     Galaxies: Seyfert --
	     Radio lines: galaxies }   
  \maketitle
%

\section{Introduction}

Multiline observations of different molecular  tracers can probe the feedback of star formation and/or nuclear activity on the chemistry and energy balance of the interstellar medium of galaxies. There is mounting observational evidence suggesting that the excitation and chemistry of the main molecular species in active galactic nuclei (AGN) are different with
respect to those found in purely star-forming galaxies (`starbursts', SBs) \citep[e.g][]{Tac94, Koh01, Use04, Mar06, Kri08, Ala11, Ala13, Ima13}.  The recent progress in the capabilities of millimeter and submillimeter telescopes has made detectable in the nearest sources a large set of molecular line transitions from different species, beyond CO line studies.  These observations  can be used to analyse the physical and chemical properties of the different  gas  components  within a  galaxy  as  well  as  its  dominant energetics.


 \begin{table*}[bt!]
\caption{\label{Tab1}The line components of the C$_2$H$(N=1-0)$ multiplet.}
\centering
\resizebox{0.75\textwidth}{!}{
\begin{tabular}{lccc}
\noalign{\smallskip} 
\hline
\hline
\noalign{\smallskip} 
Fine structure group & Hyperfine structure line & Frequency (GHz) & Line strength ($S_{\rm ij}$)	 \\
\noalign{\smallskip} 	      	      
\hline
\noalign{\smallskip} 	
Group-I & $J=3/2-1/2$, $F=1-1$ & 87.284 & 0.17  \\      
Group-I & $J=3/2-1/2$, $F=2-1$ & 87.317 & 1.67  \\  
Group-I & $J=3/2-1/2$, $F=1-0$ & 87.329 & 0.83  \\ 
Group-II & $J=1/2-1/2$, $F=1-1$ & 87.402 & 0.83  \\  
Group-II & $J=1/2-1/2$, $F=0-1$ & 87.407 & 0.33  \\  
Group-II & $J=1/2-1/2$, $F=1-0$ & 87.447 & 0.17  \\   
 \noalign{\smallskip}    
\hline
\hline 
\end{tabular}} 
\end{table*}


The C$_2$H (ethynyl)  multiplet at 3mm has a critical density of $2\times10^5$ cm$^{-3}$ at $T_{\rm K}=100~K$ (not including radiative 
trapping effects) and hence traces relatively dense gas.  However, as for other molecular transitions \citep[e.g.][]{Shi15}, the effective excitation 
density of the C$_2$H multiplet, defined as that producing emission strong enough to be observed (above a $\sim1$~K~km~s$^{-1}$ detection threshold in Shirley's study),  can be up to a factor of 10 lower. The emission of C$_2$H has been detected in a wide variety of molecular cloud environments in our Galaxy from dense Galactic cores ($n$(H$_2)\geq10^{5-6}$cm$^{-3}$) to more diffuse clouds ($n$(H$_2)\leq10^{4-5}$cm$^{-3}$).  In the early studies of Galactic molecular clouds the abundances of ethynyl, defined as $X$(C$_2$H)~$\equiv$~$N$(C$_2$H)/$N$(H$_2$), 
 were seen to increase from dense gas, where  $X$(C$_{2}$H)~$\simeq(1-60)\times10^{-10}$ \citep{Woo80, Hug84, Wat83, Wat88}, to diffuse and translucent gas clouds, where  $X$(C$_{2}$H)~$\simeq2\times10^{-8}$ \citep{Tur99}. The reason behind this trend is that in diffuse molecular gas CO can be more easily dissociated by UV photons.  This process releases carbon for C$_2$H formation and increases the abundance of this species in the gas phase, as predicted by stationary-state chemical models \citep[e.g.][]{Ste95}. 

Because of the key role of UV photons, the chemistry of C$_2$H is generally linked to ionized carbon, so that its abundance is a good tracer of the UV-pervaded outer boundary ($A_{\rm v}=1-5^{\rm m}$) of Galactic molecular clouds, known as Photon Dominated Regions (PDR), where $X$(C$_{2}$H)~$\geq10^{-8}$ \citep{Fue93, Pet05, Beu08, Wal10, Pill13, Cua15}. The changes in ethynyl abundances have also been interpreted in terms of chemical evolution in time-dependent models where chemically younger regions at lower densities would favor higher abundances of C$_2$H \citep{Li12, Pan17}. Alternatively, \citet{Bay11a, Bay11b}  showed that C$_2$H can be a tracer of regions influenced by the dissipation of turbulence and waves, heating the gas and accelerating cosmic rays. 

The abundance of  C$_2$H has also  been found to be comparable to that of molecular species like HCN or HCO$^+$ in a wide variety of external galaxies: $X$(C$_{2}$H)~$\geq10^{-9}-10^{-8}$. These extragalactic observations include SBs \citep{Hen88, Wan04, Mei05, Mar06, Ala11, Nak11, Mei12, Mei15}, AGNs \citep{Bay11b, Nak11, Ala13, Mar15}, low-metallicity objects \citep{Hei99}, (ultra) luminous infrared galaxies ((U)LIRGs) \citep{Cos11} and isolated `normal' galaxies \citep{Mar14}. Although the detection of significant  C$_2$H emission in galaxies has been generally linked to PDR chemistry in star-forming regions, other scenarios that point out more specifically to outflow-driven chemistry could provide an alternative explanation to the detection of ethynyl in the molecular outflow of Maffei~2 published by \citet{Mei12}. 

\subsection{The target: NGC~1068}
 NGC~1068 is   a   prototypical   nearby
($D\simeq14$~Mpc) Seyfert 2~galaxy that has been the subject of numerous observational  campaigns conducted with single-dish telescopes  devoted to study the fuelling of its central region and related feedback using molecular line observations  \citep[e.g][]{Use04, Isr09, Kam11, Hai12, Ala13}. NGC~1068 is an archetype of a composite SB/AGN galaxy. High-resolution observations of CO, and of more specific tracers of dense gas, such as HCN, HCO$^+$, CS, and SiO, have proved that interferometers are needed to disentangle the SB/AGN components in the circumnuclear region of this galaxy by spatially resolving the distribution, kinematics, and excitation of molecular gas \citep{Tac94, Sch00, GB10, Kri11, GB14, Tak14, Nak15, GB16, Ima16, Gal16}.

\citet{GB14} used ALMA to map the emission of a set of dense molecular gas tracers (CO(3--2), CO(6--5), HCN(4--3), HCO$^+$(4--3) and 
 CS(7--6)) in the central $r\sim2$~kpc of NGC~1068 with spatial resolutions $\simeq0.3\arcsec-0.5\arcsec$ (20--35~pc). The CO(3--2) line 
 emission in the ALMA map stems from three regions: 1) the circumnuclear disk (CND), 2) the bar, and 3) the SB ring. 
Most of the CO(3--2) flux in the ALMA map of \citet{GB14} comes from the SB ring: a two-arm spiral structure that starts from the ends of the stellar bar and forms a molecular gas pseudo-ring at $r\sim18\arcsec$($\sim1.3$~kpc). The SB ring concentrates the bulk of the massive star formation in the galaxy. The CND is an elliptical ring of 350~pc-diameter, off-centered relative to the AGN.  
In the immediate vicinity of the CND, the CO(3--2) emission is detected in an {\em arc}-like component on the north-eastern side of the disk at distances $r\simeq5-6\arcsec$(400~pc) from the AGN.  The anomalous velocities measured both in the {\em arc} feature and  the CND reveal an AGN-driven massive molecular outflow.  \citet{GB14} concluded that the molecular outflow is launched when the ionization cone of the narrow line region (NLR) sweeps the disk in the CND and further to the north in the {\em bow-shock arc}. 
More recently, ALMA observations of the CND in the CO(6--5), HCN(3--2) and HCO$^+$(3--2) lines, done with spatial resolutions $\simeq4-10$~pc, have been able to isolate and image the dust continuum and molecular line emission from a $\simeq7-10$~pc--diameter disk, which represents the submillimeter counterpart of the putative AGN torus of this galaxy \citep{GB16, Ima16, Gal16}.

Molecular line ratios analysed by \citet{GB14} and \citet{Vit14} are 
significantly different in the SB ring and the CND. The change in molecular line ratios inside
the CND  indicate  that radiative and mechanical feedback from the AGN has changed dramatically the physical conditions of molecular gas and the chemical abundances of some molecular species in the outflow region \citep[see also][]{Kel17}. The SB ring is colder and less dense than the CND, and there are also differences in their chemistry \citep{Vit14}. 

The emission of the C$_2$H multiplet in the disk of NGC~1068 was first observed by \citet{Cos11} and \citet{Ala13} with the IRAM 30m single-dish telescope. In particular, \citet{Ala13} used a chemical model to reproduce the average C$_2$H abundances derived from their data but they were not able to discern whether the C$_2$H emission was arising from PDRs, dense or shocked gas, due to the insufficient spatial resolution of these single-dish observations ($\simeq29\arcsec$).
In this work we use the unique high spatial resolution ($\leq1\arcsec$) and sensitivity capabilities of ALMA  to image the emission of C$_2$H  in the different environments of the NGC~1068 disk.
We also use a set of newly developed shock models, gas-phase PDR models and gas-grain chemical models to account for the abundances derived for C$_2$H.

 Hereafter we assume a distance to NGC~1068 of
$D\simeq14$~Mpc \citep{Bla97}; the latter implies a spatial scale of $\simeq70$~pc/$\arcsec$.

   \begin{figure*}
   \centering
    \includegraphics[width=1\textwidth]{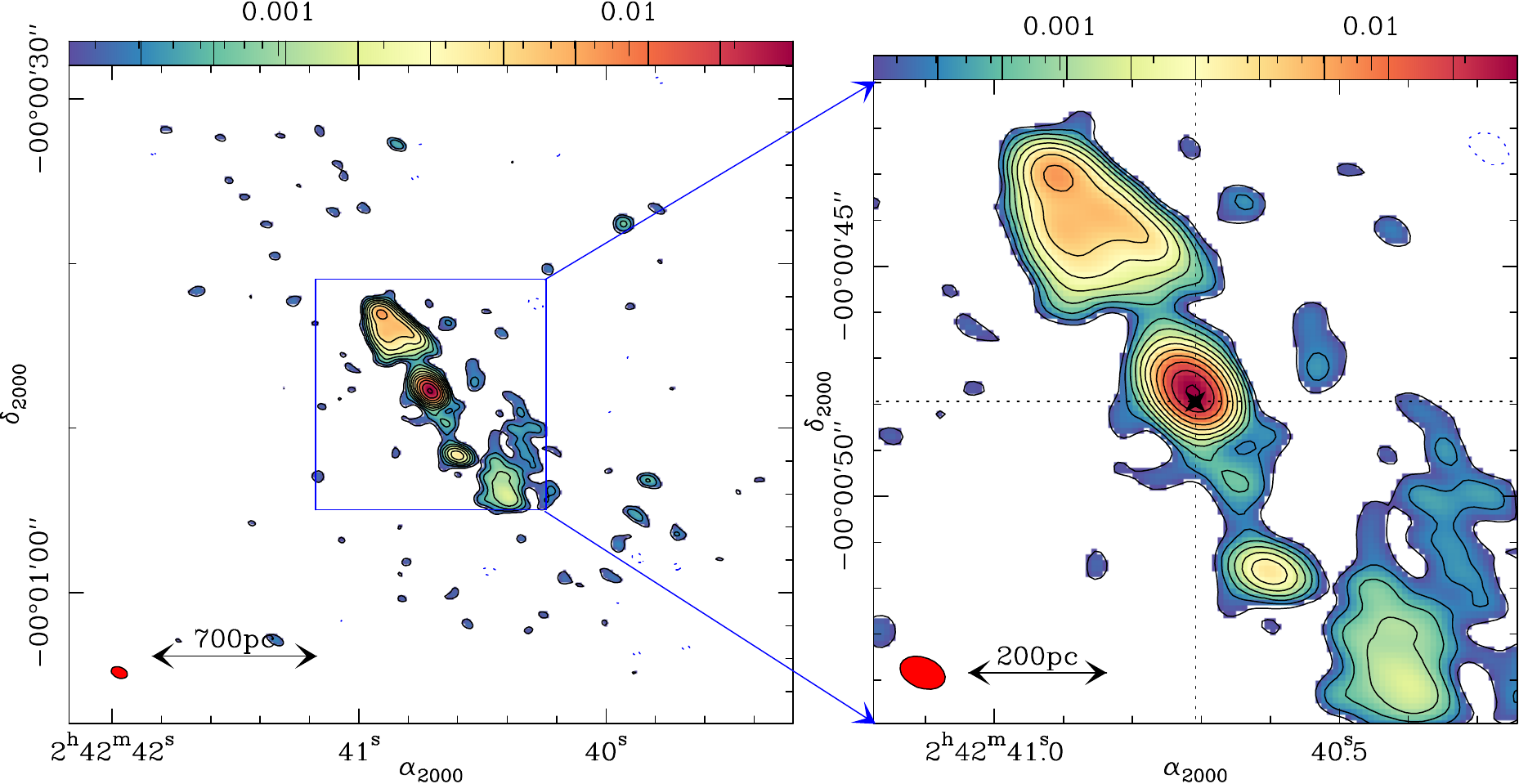}
   \caption{{\it Left panel:}~The continuum emission map of NGC~1068 obtained with ALMA at 86.3~GHz.  The map is shown in color scale with contour levels starting at -3$\sigma$ (dashed contour), and continuing with logarithmic spacing from 3$\sigma$ to 350$\sigma$ in steps of 0.21~dex, where 1$\sigma$~=~80$~\mu$Jy~beam$^{-1}$. ~{\it Right panel:}~Same as {\it left panel} but zooming in on the inner region. The position of the AGN  is highlighted by the star marker. The (red) filled
ellipses at the bottom left corners in both panels represent the beam size of ALMA at  86.3~GHz  ($1\farcs0\times0\farcs7$ at  $PA=69^{\circ}$).}
   \label{Fig1}
    \end{figure*}

\section{Observations}\label{Obs}

\subsection{ALMA data}\label{ALMA}

 We observed the emission of the hyperfine multiplet of C$_2$H($N=1-0$) in the central $r\simeq35\arcsec$(2.5~kpc)-region of
the disk of NGC~1068 with ALMA during Cycle~2 (project-ID: $\#$2013.1.00055.S; PI: S.~Garc\'{\i}a-Burillo). We used Band~3  receivers and a single pointing.  We observed one track in August 2015 with 34 antennas with projected baselines ranging from 12~m to 1430~m. 
The phase tracking center was set to $\alpha_{2000}=02^{h}42^{m}40.771^{s}$, 
  $\delta_{2000}=-00^{\circ}00^{\prime}47.84\arcsec$, which is the galaxy's center according to SIMBAD taken from the Two Micron 
  All Sky Survey--2MASS survey \citep{Skr06}. The tracking center is offset by $\leq$1$\arcsec$ relative to the AGN position:   
  $\alpha_{2000}=02^{h}42^{m}40.71^{s}$, $\delta_{2000}=-00^{\circ}00^{\prime}47.94\arcsec$ \citep{Gal96,Gal04,GB14,GB16, Gal16, Ima16}.
   The CND and the SB ring of NGC~1068 are both located inside the central 
  $r\simeq25\arcsec$(1.8~kpc) of the disk, i.e., well inside the field-of-view of the 70$\arcsec$-FWHM primary beam of the single-pointed observations. 
  
  We used four spectral windows  each with a bandwidth of 1.875~GHz and a spectral resolution of $\simeq0.98$~MHz (3.4~km~s$^{-1}$). Two of them  were placed in the lower side band (LSB) and the other two in the upper sideband (USB). The lowest frequency window of the LSB  was centered on 
the sky frequency 86.3082~GHz. This setup allowed us to observe the whole hyperfine set of lines of C$_2$H.
The C$_2$H ($N=1-0$) multiplet, hereafter referred to as C$_2$H ($1-0$), comprises six hyperfine components grouped in two fine structure groups denoted as group-I and group-II in Table~\ref{Tab1}. For the typical linewidths observed in NGC~1068 we can resolve in velocity the two fine groups but not the lines they consist of. If the opacity of the lines is low, we expect that the strongest hyperfine component  corresponds to  $J=3/2-1/2$, $F=2-1$ at  $\nu=87.317$~GHz (rest frequency). We therefore take this frequency as the reference for the velocity scale, which is hereafter derived relative to $v_{\rm sys}^{\rm HEL}$=1136~km~s$^{-1}$ \citep{GB16}.

We calibrated the data using the ALMA reduction package {\tt CASA\footnote{http//casa.nrao.edu/}}.  The calibrated uv-tables 
were transformed to {\tt GILDAS\footnote{http://www.iram.fr/IRAMFR/GILDAS}}-readable format where the mapping and cleaning 
were performed. The angular resolution obtained using natural weighting was $1\farcs0~\times~0\farcs7$ (70~pc~$\times~50$~pc) at a position angle  
$PA=69^{\circ}$ in the line and continuum data cubes.  The conversion factor between 
Jy~beam$^{-1}$ and K is 240~K~Jy$^{-1}$~beam.  The line data cube was binned to a frequency 
resolution of $\simeq2.92$~MHz ($\simeq$10~km~s$^{-1}$). The estimated 1$\sigma$-rms in the line data cube, derived in line-free 
emission areas, is  $\simeq0.4$~mJy~beam$^{-1}$ in $\simeq10$~km~s$^{-1}$-channels. An image of the continuum emission at a 
mean frequency of 86.3~GHz was obtained by averaging 60 channels ($\simeq600$~km~s$^{-1}$) free of line emission 
around the  C$_2$H group. The corresponding point source sensitivity for the continuum is 80~$\mu$Jy~beam$^{-1}$.   All the maps 
have been corrected for the attenuation by the primary beam, assuming a FWHM of 70$\arcsec$ for the latter. We estimate that the absolute flux accuracy in the maps is  about 10--15$\%$.

As our observations do not contain short-spacing correction, we expect to start filtering out  an increasing fraction of the line and 
continuum emissions on scales $>4\arcsec$; the latter corresponds to the expected largest angular scale of the emission that can be recovered in the ALMA maps used in this work.  This will likely affect the faint emission that  extends on large-scales in the interarm region. 
\citet{Cos11} measured in NGC~1068 an integrated flux for the C$_2$H multiplet of $T_{\rm a}^{*}\Delta v\simeq8\pm2$~K~km~s$^{-1}$($47\pm12$~Jy~km~s$^{-1}$) inside the  29$\arcsec$ IRAM-30m telescope beam at 86.3~GHz. This is similar to the flux estimated by \citet{Ala13}: $\simeq7$~K~km~s$^{-1}$. The corresponding flux measured by ALMA for the same aperture is $\simeq35\pm5$~Jy~km~s$^{-1}$, which is $\simeq75-85\%$ of the 30m flux estimates.
The clumpiness of the C$_2$H-emitting gas that we observe in the SB ring and the CND,  aided by the velocity structure of the emission, explains why we recover with ALMA most of the total C$_2$H flux over these regions. This scenario has 
been also tested for other molecular gas tracers mapped by ALMA in NGC~1068 \citep{GB14, Tos17}.

 \subsection{Ancillary data}\label{Anc}
 
 We retrieved the HST NICMOS (NIC3) narrow-band (F187N,
F190N) data of NGC~1068 from the Hubble Legacy Archive (HLA) in order to obtain calibrated images for the Pa$\alpha$ line emission.  
These  images  were  completely  reprocessed with a  calibration pipeline as explained in detail in Sect.~2.2 of  \citet{GB14}. The
pixel  size  of  the  HLA  images  is  $0\farcs1$ square.  

We also use the  CO(3--2) line and continuum maps  of NGC~1068 obtained with ALMA during Cycle 0 using Band~7 receivers (project-ID: $\#$2011.0.00083.S; PI: S.~Garc\'{\i}a-Burillo) and published by \citet{GB14} and \citet{Vit14}.

\section{Results}

\subsection{Continuum map}

Figure~\ref{Fig1} shows the continuum map derived at 86.3\,GHz in NGC~1068. The bulk of the continuum emission stems from a highly structured elongated jet-like feature that emanates from the nucleus and extends out to 
$r\simeq9\arcsec$ along $PA\simeq30^{\circ}$ \footnote{At larger radii we detect a few isolated spots spread mainly across the SB ring; see Fig.~\ref{Fig5}.}. The emission comprises several components that are similar to the ones imaged in the high-resolution radio continuum maps of the galaxy obtained at cm wavelengths \citep{Wil83, Ulv87, Wil87, Gal96,Gal04}.

The brightest emission corresponds to an elongated nuclear spot close to the AGN.   The nuclear spot was spatially resolved into a quadruple source in the previous subarcsecond radio continuum maps of \citet{Gal96,Gal04}:  this includes the source S1, identified as the  AGN core, and sources S2, C, and NE, which are forming the inner section of a collimated jet that emits synchrotron emission.  A similar elongated feature was identified by \citet{Ima16} in their  1.1~mm (266~GHz) continuum map obtained with ALMA.

On larger scales, we detect emission to the southwest of the nucleus from a hot spot at $r\sim4\arcsec$ and from a diffuse radio lobe spreading out from  $r\sim4\arcsec$ to  $r\sim9\arcsec$. The SW radio lobe is  bent to the W relative to the overall axis of the whole jet-like structure ($PA\simeq30^{\circ}$).  This distortion of the jet was interpreted by \citet{Wil82} as the signature of the ram pressure of the rotating interstellar gas of the galaxy on the radio plasma. We also detect emission to the northeast of the nucleus from a radio lobe that exhibits a conical limb-brightened morphology (Fig.~\ref{Fig1}). The VLA maps at 2cm of \citet{Wil87} were the first to show this morphology for the northeast radio lobe, which was interpreted as the signature of  a bow-shock wave driven into the interstellar material as it is compressed by the radio ejecta. The ALMA maps of \citet{GB14}  showed 
an arc of gas and dust emission located at the edges of the northeast radio lobe. 
This gas feature, referred to as the
bow-shock {\it arc}, has  anomalous  outward velocities interpreted  as the result of the expansion of an AGN-driven molecular outflow inside the disk of the galaxy \citep{GB14}.

   \begin{figure*}[tbh!]
      \centering
    \includegraphics[width=0.95\textwidth]{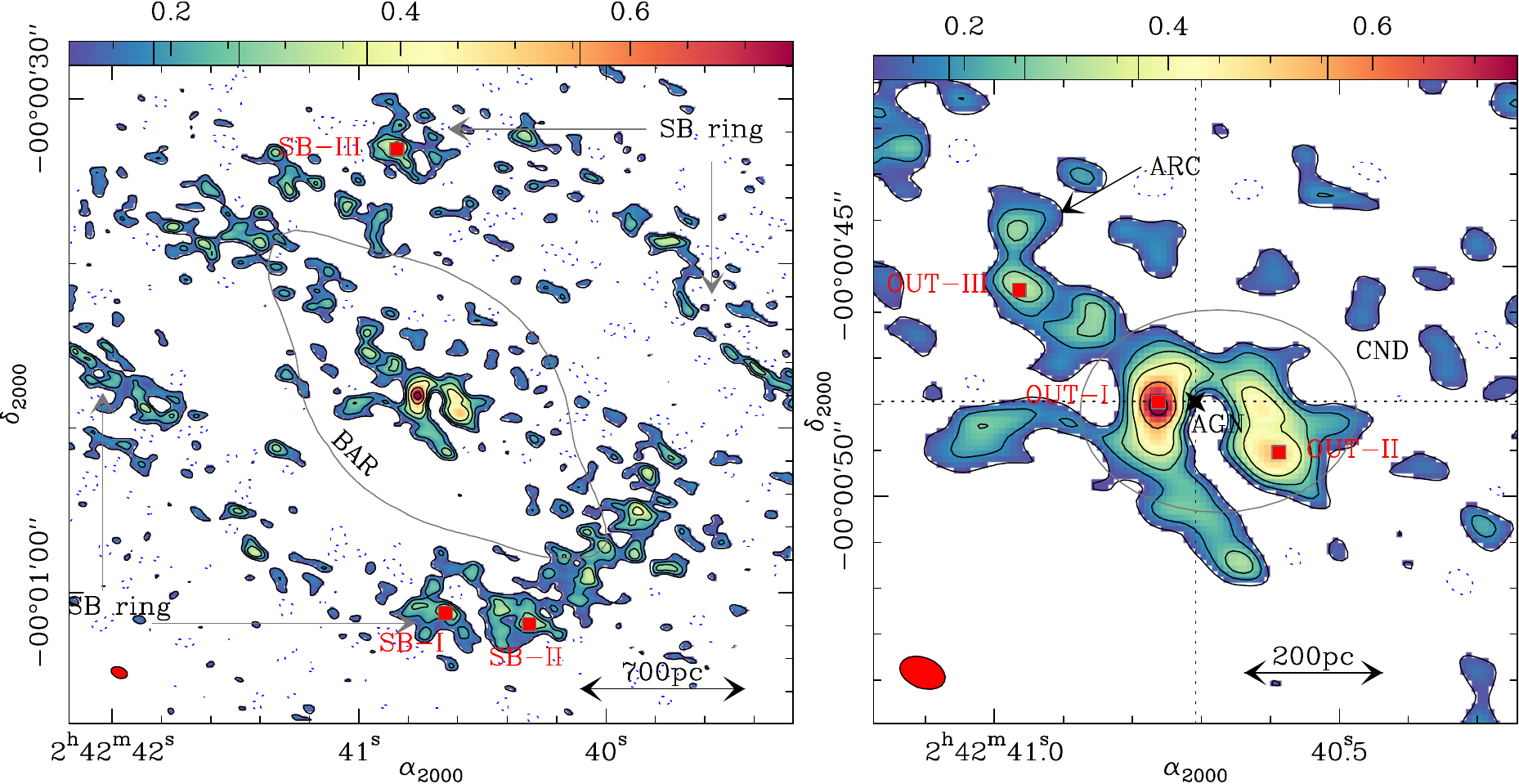}
   \caption{{\it Left panel:}~The C$_2$H(1--0) integrated intensity map obtained with ALMA in the disk of NGC~1068. The map is shown in 
color scale with contour levels  -3$\sigma$ (dashed contour), 3$\sigma$, 5$\sigma$, 7$\sigma$,10$\sigma$, 15$\sigma$, and 20$\sigma$, 
where 1$\sigma=0.037$~Jy~km~s$^{-1}$beam$^{-1}$. We identify the location of the SB ring and the bar. The latter is identified by a 
representative isophote of the NIR K-band image of 2MASS~\citep{Skr06}. The (red) square markers highlight the positions of the knots SB-I, SB-II 
and  SB-III along the SB ring.~{\it Right panel:}~Same as {\it left panel} but zooming in on the outflow region, which extends out to $r
\simeq6\arcsec$~(400~pc), as  described by \citet{GB14}. The outflow region comprises the CND and the bow-shock arc.  The position of the 
AGN ($\alpha_{2000}=02^{h}42^{m}40.71^{s}$, $\delta_{2000}=-00^{\circ}00^{\prime}47.94\arcsec$) is highlighted by the star marker.  The (red) 
square markers highlight the positions of the knots OUT-I, OUT-II and OUT-III. The (red) filled
ellipses at the bottom left corners in both panels represent the C$_2$H beam size ($1\farcs0\times0\farcs7$ at  $PA=69^{\circ}$).}
   \label{Fig2}
    \end{figure*}

   \begin{figure*}[bth!]
      \centering
    \includegraphics[width=0.73\textwidth]{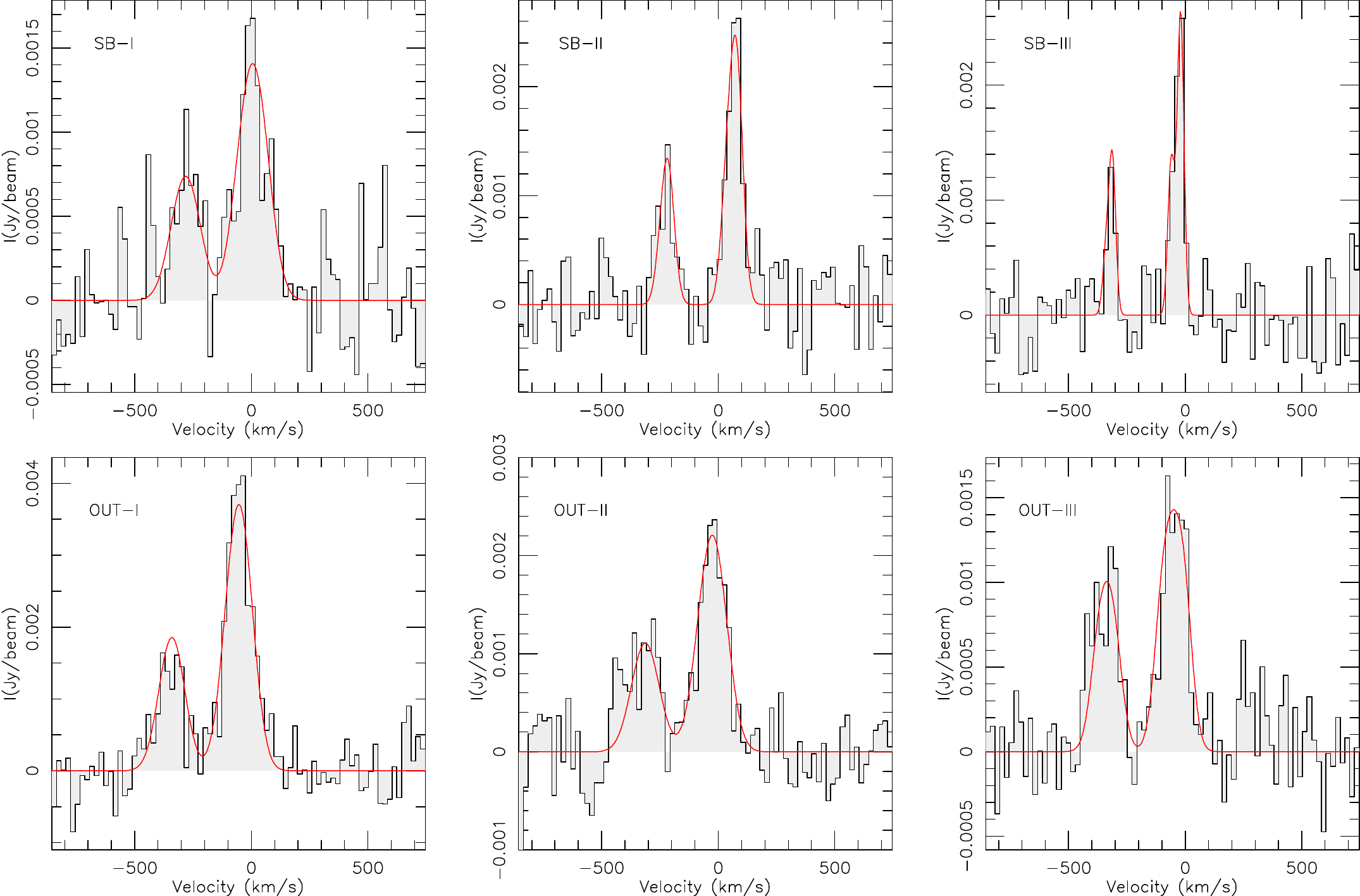}
   \caption{We show in the {\it upper panels} the C$_2$H emission spectra taken at three representative knots of the SB ring (SB-I, SB-II and SB-III).  The {\it lower panels} display the C$_2$H spectra taken at three representative knots of the outflow region (OUT-I, OUT-II and OUT-III). The apertures used to extract the spectra correspond to the beam size at the location of each knot ($1\farcs0\times0\farcs7$ at  $PA=69^{\circ}$). The positions of the knots are defined in Fig.~\ref{Fig2}. The velocity scale for each spectrum has been derived for the strongest C$_2$H line at 87.317~GHz (rest frequency) and is relative to $v_{\rm sys}^{\rm HEL}$=1136~km~s$^{-1}$. The (red) curves show the composite Gaussian profiles of the best-fit to the emission, as described is Sect.~\ref{opt}.}
   \label{Fig3}
    \end{figure*}

\subsection{C$_2$H map}\label{C2H-map}

Figure~\ref{Fig2} shows the C$_2$H(1--0) velocity-integrated intensity map obtained with ALMA in the disk of NGC~1068. Intensities have been integrated at each pixel inside the velocity interval defined by: 
$v-v_{\rm sys}^{\rm HEL}\subset[ -510, 200]$~km~s$^{-1}$. This window encompasses any potential  emission stemming from the four strongest hyperfine components of C$_2$H(1--0), which are expected to arise at $v-v_{\rm sys}^{\rm HEL}$~$\simeq$~0, --40, --290, and --310~km~s$^{-1}$. When defining the velocity window we further assume that the disk rotation of NGC~1068 can cause
a maximum velocity shift of the line of $\simeq\pm200$~km~s$^{-1}$  around $v_{\rm sys}^{\rm HEL}$  \citep{Sch00, Kri11, GB10, GB14}.

Figure~\ref{Fig3} shows the  C$_2$H(1--0)  spectra extracted from a representative set of positions that correspond to strong ($>7\sigma$ in velocity-integrated units) emission knots in the outflow (OUT-I-to-III) and the SB region (SB-I-to-III).   Due to the expected non-negligible velocity gradient inside the ALMA beam caused by rotation, non-circular motions and turbulence, the individual hyperfine components of the fine groups are blended in all the spectra shown in Fig.~\ref{Fig3}. We nevertheless resolve in the line profiles the emission from the two fine structure groups, which are shifted to each other by $\simeq300$~km~s$^{-1}$. Hereafter we define the {\em main} and the {\em secondary} group of lines as the two strongest pairs of hyperfine lines that appear blended around   $<v-v_{\rm sys}>^{\rm HEL}~\simeq$~--20 in group-I,  and --300~km~s$^{-1}$ ($\pm 200$~km~s$^{-1}$) in group-II, respectively (see Table\ref{Tab1}). 
 No significant emission is detected from the two weakest hyperfine components  of C$_2$H in the individual spectra, i.e., outside the range defined by the {\em main} and {\em secondary} groups. Emission from these {\em satellite} lines should lie at $v-v_{\rm sys}^{\rm HEL}$~$\simeq$~+110 and --450~km~s$^{-1}$ ($\pm 200$~km~s$^{-1}$)  and likely have intensities a factor of $\sim10$ lower than the strongest line of the C$_2$H multiplet, unless the lines were mostly optically thick, a scenario that we discard in Sect.~\ref{opt}. Therefore we conclude that the interval defined above to derive the velocity-integrated intensities of the C$_2$H group of lines is not missing any significant emission.

\begin{figure*}[bht!]
      \centering
    \includegraphics[width=1\textwidth]{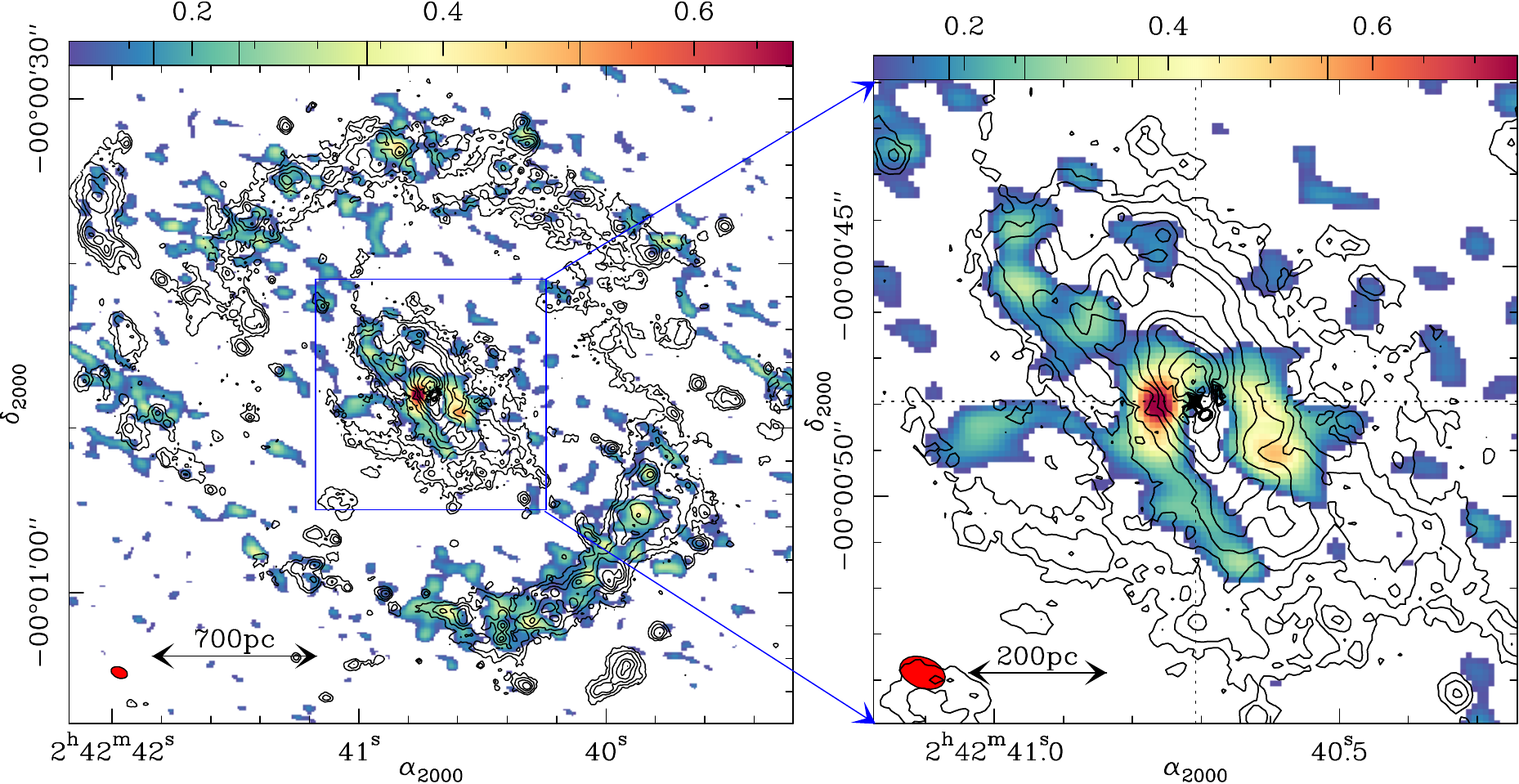}
   \caption{{\it Left panel:}~Overlay of the Pa$\alpha$ emission HST map (contours) on the C$_2$H(1--0) integrated intensity map (color scale as in Fig.~\ref{Fig2}). Contours of Pa$\alpha$  emission are displayed with logarithmic spacing from 2.4 to 1600 counts~s$^{-1}$pixel$^{-1}$ in steps of 0.28~dex .~{\it Right panel:}~Same as {\it left panel} but zooming in on the outflow region. The position of the AGN  is highlighted by the star marker. The (red) filled ellipses at the bottom left corners in both panels represent the C$_2$H beam size.}
   \label{Fig4}
    \end{figure*}

As shown by Fig.~\ref{Fig2}, a sizeable fraction  ($\simeq65\%$) of the total (spatially-integrated) C$_2$H line flux recovered by ALMA stems from the SB ring region.   Figure~\ref{Fig4} shows the overlay of the Pa$\alpha$ HST map on the C$_2$H ALMA map. The SB ring concentrates the bulk of the recent massive star formation in the disk traced by the Pa$\alpha$ emission complexes associated with H{\small II} regions. 
 The emission of C$_2$H is mostly prominent at the south-western end of the SB ring, a region hosting an intense star formation activity characterized  by  strong Pa$\alpha$ emission. The molecular line emission in the SB ring is very clumpy and it comes from cloud complexes of $\geq100$~pc-size. 
 
However,  the brightest C$_2$H emission originates from  the CND. The morphology of the CND displayed by other molecular tracers 
imaged by ALMA \citep{GB14, Tak14, Tos17} is to a large extent echoed by C$_2$H: the CND appears as an elongated
ring  with two spatially-resolved knots  of emission asymmetrically located east and west of the AGN locus (denoted as knots OUT-I and 
OUT-II, respectively; see Fig.~\ref{Fig2}). The east and west knots  are bridged by fainter emission north of the AGN. We also detect 
significant emission that connects the CND with the outer disk:  in the bow-shock {\em arc} region (including knot OUT-III; see Fig.~\ref{Fig3}) 
to the northeast, and in two emission lanes east and south of the CND.

\begin{figure*}
   \centering
    \includegraphics[width=0.95\textwidth]{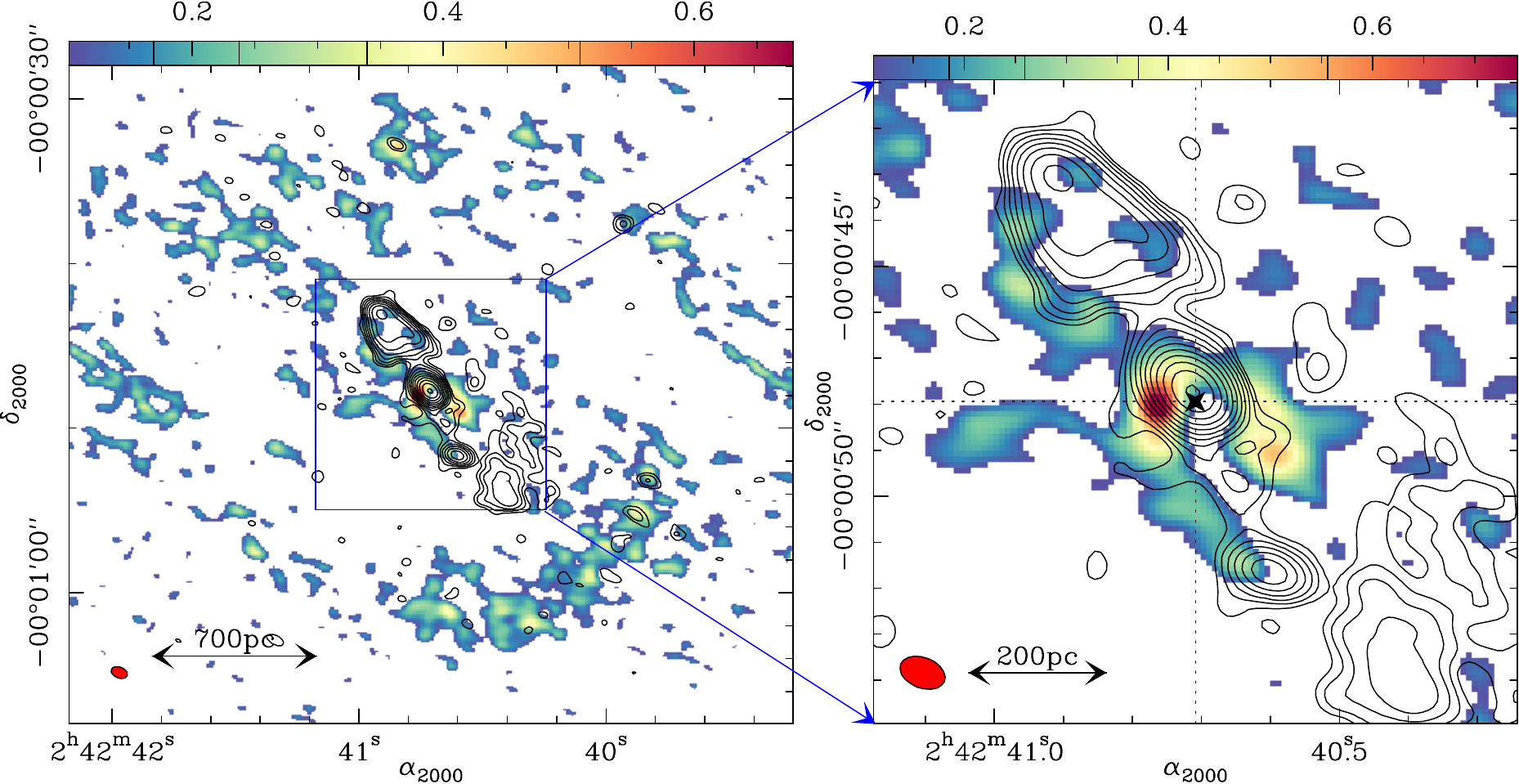}
   \caption{{\it Left panel:}~Overlay of the continuum emission map of NGC~1068 obtained with ALMA at 86.3~GHz (contours) on the C$_2$H(1--0) integrated intensity map (color scale as in Fig.~\ref{Fig2}). Contours of continuum emission are displayed with logarithmic spacing from 3$\sigma$ to 350$\sigma$ in steps of 0.21~dex, where 1$\sigma$~=~80$~\mu$Jy~beam$^{-1}$. ~{\it Right panel:}~Same as {\it left panel} but zooming in on the outflow region. The position of the AGN  is highlighted by the star marker. The (red) filled
ellipses at the bottom left corners in both panels represent the C$_2$H beam size.}
   \label{Fig5}
    \end{figure*}

Figures~\ref{Fig4} and~\ref{Fig5} respectively show in contours the Pa$\alpha$ and the continuum emission (same as in Fig.\ref{Fig1}) on the C$_2$H ALMA map, in color. Figures ~\ref{Fig4} and \ref{Fig5} evidence that the C$_2$H emission lies at the edges of the AGN ionized nebulosity  traced by Pa$\alpha$
 and radio continuum emission. \citet{GB14} analysed the morphology and kinematics of molecular gas in this region, traced by the CO(3--2) line, and concluded that the NLR gas within the ionization cone  is sweeping the ISM of the disk creating  a coplanar molecular outflow out to a radius of 400~pc, i.e., a region that comprises the CND and the bow-shock {\em arc} feature. The C$_2$H line seems to probe better than CO the interface between the molecular disk and the ionized gas outflow on the same spatial scales. Hereafter we will thus refer to the region encompassing the CND and the bow-shock {\em arc} as the {\em outflow} region.
 
\begin{figure*}[tb!]
   \centering
    \includegraphics[width=.95\textwidth]{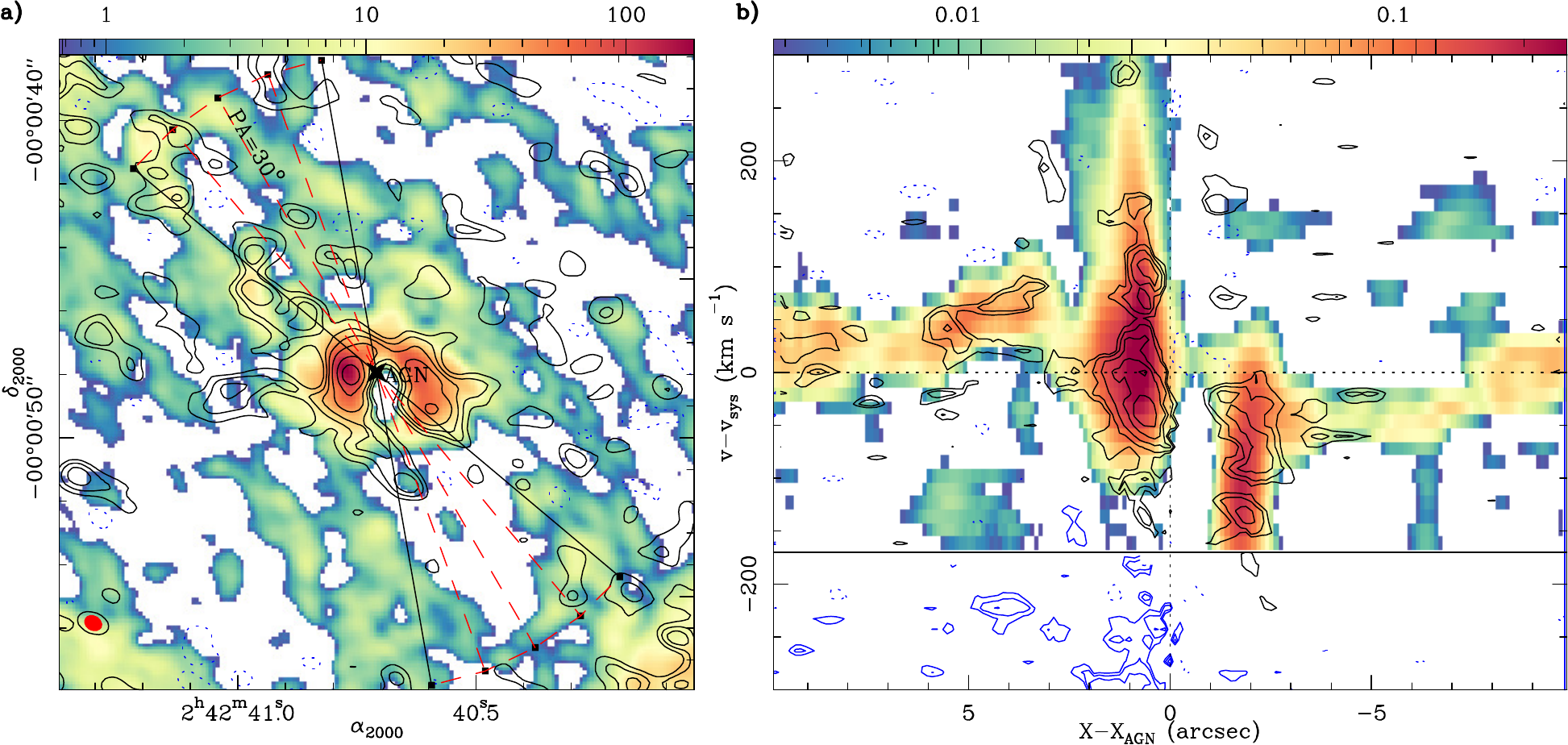}
   \caption{{\bf a)}~{\it Left panel:}~Overlay of the C$_2$H(1--0) integrated intensity map (with contours as in Fig.~\ref{Fig2}) on the CO(3--2) integrated intensity map (with color scale expanding the range [$3\sigma$, $800\sigma$], with $1\sigma=0.22$~Jy~km~s$^{-1}$) of \citet{GB14} in the central $r=10\arcsec$(700~pc) of NGC~1068. The (black) empty and (red) filled elllipses at the bottom left corner represent the C$_2$H and CO beam sizes respectively. ~{\bf b)}~{\it Right panel:}~Comparison of the {\rm average} position-velocity (p-v) plot obtained for the outflow region in C$_2$H(1--0) (black contours: -$2.5\sigma$ (dashed contour), $2.5\sigma$,  $3.5\sigma$, $5\sigma$ to $9\sigma$ by  $2\sigma$, with $1\sigma=0.18$~mJy) and CO(3--2) (color scale: [$2\sigma$, $200\sigma$], with  $1\sigma=1.3$~mJy). The averaging has been done using the axes oriented along $PA=$~10$^{\circ}$, 20$^{\circ}$, 30$^{\circ}$, 40$^{\circ}$, and 50$^{\circ}$ (shown in {\bf a)}). The average outflow axis lies at  $PA=$~30$^{\circ}$. Blue contours represent
the emission stemming from the secondary fine group of the C$_2$H(1--0) multiplet (group-II).  
   Offsets along the x axis are measured in arc seconds relative to the AGN locus. Velocity scale is measured for the strongest C$_2$H line at 87.317~GHz (rest frequency) and is relative to $v_{\rm sys}^{\rm HEL}$=1136~km~s$^{-1}$. The black horizontal line indicates the edges of the CO(3--2) band. }
   \label{Fig6}
    \end{figure*}

Figures~\ref{Fig6}a and b compare, respectively, the morphology and the kinematics of molecular gas in the outflow region derived from the C$_2$H(1--0) and CO(3--2) images obtained by ALMA. The distribution of C$_2$H shows a good correspondence with the brightest emission clumps of CO in the outflow, both in the CND and in the bow-shock {\em arc}, although the emission of CO is comparatively  more widespread than that of C$_2$H in this region (see Fig.~\ref{Fig6}a). Figure~\ref{Fig6}b compares the emission of C$_2$H and CO in the outflow using the position-velocity (p-v)  parameter space. With the objective of increasing the signal to noise ratio, more critical for  C$_2$H, the outflow p-v plot diagrams shown in Fig.~\ref{Fig6}b have been derived by averaging the emission along the five axes shown in Fig.~\ref{Fig6}a, defined by $PA=$~10$^{\circ}$, 20$^{\circ}$, 30$^{\circ}$, 40$^{\circ}$, and 50$^{\circ}$. These axes cover a sizeable fraction of the emission along the molecular outflow, which is oriented along $PA=$~30$^{\circ}$ \citep{GB14}. Furthermore, with the aim of isolating the non-circular motions associated with the outflow, we have subtracted the rotation curve model of \citet{GB14} projected along each individual p-v plot before averaging.
As illustrated by Fig.~\ref{Fig6}b, the emission of C$_2$H follows closely the velocities associated with the CO outflow in this region: a  significant fraction of the emission of both molecular tracers lies outside the expected range of virial  motions attributable to rotation and  dispersion: emission is up to $\geq50-200$~km/s-redshifted (blueshifted) on the northern (southern) side of the outflow out to $r\simeq6\arcsec$ (400~pc). This reflects the sign and the right order of magnitude of the mean-velocity field deviations seen in the CO(3--2) map of \citet{GB14}.

\subsection{C$_2$H line opacities}\label{opt}

We have performed a fit of the spectra shown in Fig.~\ref{Fig3} using a comb of four Gaussian profiles that correspond to the strongest hyperfine components of the C$_2$H(1--0) multiplet. Among the parameters to fit we assumed common widths and excitation temperatures, while optical depth ratios  are based on the theoretical line strengths \citep{Tuc74}. The velocity centroids and the optical depths of the strongest component ($\tau_{\rm str}$) were taken as free parameters. Table~\ref{Tab2} lists the results obtained in the fit of the spectra shown in Fig.~\ref{Fig3}. The fit provides the line parameters for the strongest 
line ($J=3/2-1/2$, $F=2-1$). We have also derived  the average  {\em secondary}-to-{\em main} velocity-integrated intensity ratios ($R$) for the CND, the bow-shock arc and the SB ring. The optical depths derived from the fits of the two spectra located in the CND (OUT-I and OUT-II) are low: $\tau_{\rm str}\leq0.1~(\pm0.05)$. This is in agreement with  the average $R$ ratio derived for the CND, $<R>~\simeq0.49\pm0.05$, which is very close to the optically thin limit for $R$ ($\simeq0.46$).  The optical depth derived for the knot  OUT-III is nevertheless higher $\tau_{\rm str}\simeq1.3~(\pm0.6)$; this result is in line with the higher average $R$ ratio measured in the bow-shock arc: $<R>~\simeq0.70\pm0.09$. 
The optical depths derived for the SB knots (SB-I, SB-II and SB-III) are low: $\tau_{\rm str}\leq0.1-0.2~(\pm0.1)$. The corresponding average $R$ ratio over the SB ring is also low  $<R>~\simeq0.56\pm0.06$. 

We can therefore conclude that  the lines of the C$_2$H(1--0) multiplet are mostly optically thin throughout the disk of NGC~1068 with the exception of the bow-shock arc region where opacities may be close to unity\footnote{In this scenario of low opacities the intensities of the weakest {\em satellite} lines would be at most at a level $\leq$2$\sigma$ at the position of the strongest emission knot OUT-I.}.

  \begin{table*}[bt!]
\caption{\label{Tab2}Parameters of C$_2$H line fitting results.}
\centering
\resizebox{0.85\textwidth}{!}{
\begin{tabular}{lcccrrc}
\noalign{\smallskip} 
\hline
\hline
\noalign{\smallskip} 
KNOT & $\alpha_{\rm 2000}$ & $\delta_{\rm 2000}$ &  $\eta_{\rm f}\times$(T$_{\rm ex}$-T$_{\rm bg})\times\tau_{\rm str}$ &  (v-v$_{\rm sys}$)$_{\rm str}$	& FHWM$_{\rm str}$	& $\tau_{\rm str}$ 	 \\
\noalign{\smallskip}
	    & $^h$ ~~~~~   $^m$ ~~~~ $^s$ &  $^o$~~~~  $'$ ~~~~  $\arcsec$ &  (K) &	(km~s$^{-1}$)	&	(km~s$^{-1}$)	    &	--  \\
\noalign{\smallskip} 	      	      
\hline
\noalign{\smallskip} 	
SB-I & 02$^h$:42$^m$:40.65$^s$ & $-00^{\circ}01^{\prime}01.2\arcsec$ & $(2.8\pm1.0)\times10^{-1}$  &  $19\pm8$ &   $139\pm22$   &   $0.2\pm0.6$ \\      
SB-II & 02$^h$:42$^m$:40.31$^s$ & $-00^{\circ}01^{\prime}01.9\arcsec$ & $(5.3\pm0.7)\times10^{-1}$  &  $79\pm3$ &   $63\pm11$   &   $0.1\pm0.2$ \\
SB-III & 02$^h$:42$^m$:40.85$^s$ & $-00^{\circ}00^{\prime}33.1\arcsec$ & $(6.6\pm0.5)\times10^{-1}$ &  $-20\pm2$ &   $34\pm10$   &   $0.1\pm0.3$ \\
OUT-I & 02$^h$:42$^m$:40.76$^s$ & $-00^{\circ}00^{\prime}48.0\arcsec$ & $(6.8\pm0.3)\times10^{-1}$    & $-41\pm3$ &   $126\pm10$   &   $0.1\pm0.1$ \\ 
OUT-II & 02$^h$:42$^m$:40.59$^s$ & $-00^{\circ}00^{\prime}49.0\arcsec$ &  $(4.1\pm1.1)\times10^{-1}$  &  $-13\pm5$ &   $135\pm16$   &   $0.1\pm1.9$ \\
OUT-III & 02$^h$:42$^m$:40.96$^s$ & $-00^{\circ}00^{\prime}45.5\arcsec$ & $(5.4\pm1.6)\times10^{-1}$  &  $-37\pm5$ &   $101\pm12$   &   $1.3\pm0.7$ \\    
 \noalign{\smallskip}    
\hline
\hline 
\end{tabular}} 
\tablefoot{All the line parameters correspond to the strongest C$_2$H($1-0$) line ($J=3/2-1/2$, $F=2-1$) and are derived according to the procedure described in Sect.~\ref{opt}: T$_{\rm ex}$ is the excitation temperature, T$_{\rm bg}$ is the background temperature ($\simeq$~2.73~K), $\eta_f$ is the unknown filling factor of the 
emission, ($v-v_{\rm sys}$)$_{\rm str}$ is the velocity centroid of the line relative to $v_{\rm sys}^{\rm HEL}$=1136~km~s$^{-1}$, FHWM$_{\rm str}$ is the half width 
at half maximum of the line, and $\tau_{\rm str}$ is the estimated peak opacity.  }\\
\end{table*}

\subsection{C$_2$H abundances} \label{abun}

In this section we estimate the abundances of C$_2$H relative to H$_2$, defined as $X$(C$_2$H)~$\equiv$~$N$(C$_2$H)/$N$(H$_2$), both in the SB ring and in the outflow region of NGC~1068. 

To estimate the total column densities of C$_2$H,  $N$(C$_2$H),  we will assume  local thermodynamic equilibrium (LTE) conditions for the sake of simplicity.  Under the assumption of LTE conditions,  the  C$_2$H
column densities can be obtained straightforwardly from the intensities of the line. As  the velocity-integrated intensities of the C$_2$H multiplet derived in Sect.~\ref{C2H-map} are a blend of four lines, we have first re-normalized the observed intensities by a factor $\simeq0.46$ defined by the expected ratio of intensities of the strongest main line ($J=3/2-1/2$, $F=2-1$) to the whole multiplet in the optically thin limit \footnote{In this scenario the factor is derived from the ratio $\simeq 1/(1+0.5+0.5+0.2)\simeq$~0.46}. The validity of the hypothesis of low optical depths for C$_2$H, central to the significance of the LTE estimates, has been discussed in Sect.~\ref{opt}. We  also need to assume a value for the  excitation temperatures ($T_{\rm ex}$) for the  C$_2$H emitting gas. As we only have one line multiplet of C$_2$H, we will assume a range of temperatures ($T_{\rm ex}$=10--150~K) that encompasses the total span  of kinetic temperatures found by \citet{Vit14} in their Large Velocity Gradient (LVG) fit of the multiline observations 
obtained for CO, HCN and HCO$^+$ in the different environments of NGC~1068.

In LTE at temperature $T_{\rm ex}$ the column density of the upper level $u$ of a particular transition is related to the total column density of the species via the Boltzmann equation:

\begin{equation}
N=\frac{N_u Z}{g_ue^{-E_u/kT_{ex}}},
\label{eq:N}
\end{equation}
where $N$ is the total column density of the species, $Z$
is the partition function evaluated for  $T_{\rm ex}$, $g_u$ is the statistical weight of the level $u$ and $E_u$ is its energy above the ground state. If the emission is assumed to be optically thin, and if we adopt a filling factor of unity, the column density $N_u$ is related to the observed line intensity:
\begin{equation}
N_u = \frac{8\pi k \nu^2 I}{hc^3A_{ul}}
\label{eq:Nu}
\end{equation}
where $I$ is the velocity-integrated line intensity (in K km s$^{-1}$) and A$_{ul}$ is the Einstein coefficient that correspond to a particular transition, which in our case corresponds to the $J=3/2-1/2$, $F=2-1$  line of C$_2$H.


\begin{table}[tbh]
\begin{center}
\caption[]{Fractional abundances of C$_2$H in the SB-ring and the outflow regions.}
\resizebox{0.45\textwidth}{!}{
\begin{tabular}{lcc}
\hline 
\hline
\noalign{\smallskip}
 & SB-ring  & outflow   \\
 \noalign{\smallskip}
 \hline
 \noalign{\smallskip}
 T$_{\rm ex}$=10 K & $(1.3\pm0.7)\times10^{-8}$ & $(1.6\pm0.8)\times10^{-8}$  \\
 T$_{\rm ex}$=25 K & $(9.7\pm4.0)\times10^{-8}$ & $(1.1\pm0.4)\times10^{-7}$  \\
T$_{\rm ex}$=50 K & $(4.3\pm2.0)\times10^{-7}$ & $(4.8\pm2.0)\times10^{-7}$  \\
T$_{\rm ex}$=100 K & $(1.8\pm0.8)\times10^{-6}$ & $(2.0\pm0.8)\times10^{-6}$  \\
T$_{\rm ex}$=150 K & $(4.2\pm2.0)\times10^{-6}$ & $(4.7\pm2.0)\times10^{-6}$  \\
\noalign{\smallskip}
\hline
\hline
\end{tabular}\label{Tab3}}
\end{center}
\tablefoot{Abundances are estimated from the C$_2$H ($1-0$) line and using gas column densities derived from the dust continuum emission at  349~GHz. We assume LTE conditions and a range of excitation temperatures as explained in Sect.~\ref{abun}}\\
\end{table}


Furthermore, in order to derive $X$(C$_2$H) we have estimated  $N$(H$_2$) from the ALMA dust continuum emission map of the galaxy obtained at 349~GHz by \citet{GB14}. These observations well probe the molecular gas column densities for the regions studied here:  first, because the dust continuum emission at 349~GHz is mostly optically thin, and, second, because the contribution from atomic hydrogen to the total neutral gas content is negligible in the central $r\simeq30\arcsec\simeq2$~kpc of the disk \citep{Bri97, GB14}. 
We have used a modified black-body law to derive the dust mass ($M_{\rm dust}$). In this approach, the fluxes measured in a given aperture, $S_{\rm 349~GHz}$, can be expressed as $S_{\rm 349~GHz}=M_{\rm dust}\times\kappa_{\rm 349~GHz}\times B_{\rm 349~GHz}$($T_{\rm dust}$)/$D^{2}$, where the emissivity of dust at 349~GHz,  
$\kappa_{\rm 349~GHz}\simeq\kappa_{{\rm 352~GHz}}\simeq0.09$~m$^2$~kg$^{-1}$\citep{Kla01}, $B_{\rm \nu}$($T_{\rm dust}$) is the Planck function, and $D$ is the distance.

 The value of $M_{\rm dust}$ per beam was used to predict the associated molecular gas column density map. To this aim,  we applied the linear dust/gas scaling ratio of \citet{Dra07} to the gas-phase oxygen abundances measured in the central 2~kpc of NGC~1068 \citep[$\sim 12 + $log~(O/H)~$\sim8.8$;][]{Pil04, Pil07}, which yields a gas--to--dust mass ratio of $\sim60^{+30}_{-30}$~\citep[see discussion in][]{GB14}. Finally, we have degraded the spatial resolution of the resulting column density map to match that of the C$_2$H map prior to obtaining a map for $X$(C$_2$H).  
 
 Table~\ref{Tab3} lists the mean fractional abundances of C$_2$H as estimated by our LTE analysis for the reported range of $T_{\rm ex}$. In the interest of exploring  any potential difference between  the SB and the outflow, we have spatially averaged $X$(C$_2$H) inside these two regions, as defined in Sect.~\ref{C2H-map}. As expected, the values of $X$(C$_2$H) depend heavily on the assumed $T_{\rm ex}$. For a given $T_{\rm ex}$, the abundance of  C$_2$H  is marginally $\simeq10-20\%$ higher in the outflow region than in the SB ring, with a common lower limit of   $X$(C$_2$H) $\simeq$ a few 10$^{-8}$ for the lowest temperature we assumed: $T_{\rm ex}$~=~10~K. We note that this is close to the formal lower limit on $T_{\rm ex}\geq8$~K   that is derived from the values fitted for $\eta_{\rm f}\times$(T$_{\rm ex}$-T$_{\rm bg})\times\tau_{\rm str}$, listed in Table~\ref{Tab3}. 
This lower limit is formally obtained by assuming a very conservative upper limit on the filling factors  $\eta_{\rm f}\leq1$. All in all, the reported high abundances of C$_2$H found in the SB ring and the outflow for  $T_{\rm ex}$~=~10~K are both comparable to those found in Galactic and extragalactic PDRs associated with massive star formation. However, although this scenario seems plausible in the SB ring, where molecular gas is currently feeding a massive star formation episode in NGC~1068, the evidence of a similar level of star formation activity that may sustain these high abundances of C$_2$H in the molecular  outflow is less clear. 
This suggests that a  different mechanism is at play in this region.


 \begin{table}
 \caption{Grid of PDR models (UCL\_PDR).}
\centering
\resizebox{0.32\textwidth}{!}{
\begin{tabular}{cccc}
\hline
\hline
\noalign{\smallskip}
Model & $\zeta$ & $\chi$ &  n$_{\rm H}$   \\
\noalign{\smallskip}
&  ($\zeta_0) $ & (Draine field) & (cm$^{-3}$) \\
\noalign{\smallskip}
\hline
\noalign{\smallskip}
48& 1.0 &10 &  10$^3$  \\ 
49& 1.0 &100 &  10$^3$  \\ 
50& 1.0 &1000 &  10$^3$  \\ 
51& 10 & 100 &  10$^3$  \\ 
\noalign{\smallskip}
\hline
\hline
\end{tabular}\label{Tab4}}
\tablefoot{Column (1) lists the Model number. Columns (2) and (3) list, respectively, the cosmic-ray ionization rates ($\zeta$) and radiation fields ($\chi$) in units of their standard Galactic values ($\zeta_o$ = 5~$\times$~10$^{-17}$ s$^{-1}$ and 1 Draine field~$\equiv$~2.74~$\times$~10$^{-3}$~erg~s$^{-1}$~cm$^{-2}$). Column (4) lists the hydrogen gas density (n$_{\rm H}$). For the PDR models the temperature varies with depth and is not listed.}\\
\end{table}
 

 
 \begin{table}
\caption{Grid of gas-grain chemical models (UCL\_CHEM). }
\centering
\resizebox{0.48\textwidth}{!}{
\begin{tabular}{cccccc}
\hline
\hline
\noalign{\smallskip}
Model & $\zeta$ & $\chi$ & T$_{\rm k}$ & n$_{\rm H}$ & A$_V$ \\
\noalign{\smallskip}
 & ($\zeta_0) $ & (Draine field) & (K) & (cm$^{-3}) $ & (mag)  \\
\noalign{\smallskip} 
\hline 
\noalign{\smallskip}
 1&   1.0&  1.0&100.0&     10$^4$& 1\\
 2&   1.0&  1.0&100.0&     10$^4$&50\\
 3&   1.0&  1.0&100.0&    10$^5$& 1\\
 4&   1.0&  1.0&100.0&    10$^5$&50\\
 5&   1.0&  1.0&100.0&   10$^6$&50\\
 6&  10.0&  1.0&100.0&     10$^4$& 1\\
 7&  10.0&  1.0&100.0&     10$^4$&50\\
 8&  10.0&  1.0&100.0&    10$^5$& 1\\
 9&  10.0&  1.0&100.0&    10$^5$&50\\
10&  10.0&  1.0&100.0&   10$^6$&50\\
11&   1.0& 10.0&100.0&     10$^4$& 1\\
12&   1.0& 10.0&100.0&     10$^4$&50\\
13&   1.0& 10.0&100.0&    10$^5$& 1\\ 
14&   1.0& 10.0&100.0&    10$^5$&50\\
15&   1.0& 10.0&100.0&   10$^6$&50\\
16&  10.0& 10.0&100.0&     10$^4$& 1\\
17&  10.0& 10.0&100.0&     10$^4$&50\\
18&   1.0&500.0&100.0&     10$^4$& 1\\
19&   1.0&500.0&100.0&     10$^4$&50\\
20& 500.0&  1.0&100.0&     10$^4$& 1\\
21& 500.0&  1.0&100.0&     10$^4$&50\\
22&5000.0&  1.0&100.0&     10$^4$& 1\\
23&5000.0&  1.0&100.0&     10$^4$&10\\ 
24&   0.0&  1.0&100.0&    10$^5$& 1\\
25&   0.0&  1.0&100.0&    10$^5$&50\\
26&   1.0&  1.0&200.0&    10$^5$& 1\\
27&   1.0&  1.0&200.0&    10$^5$&50\\
28&   1.0&  1.0&200.0&     10$^4$& 1\\
29&   1.0&  1.0&200.0&     10$^4$&50\\
30&   1.0&  1.0&200.0&   10$^6$&50\\
31&  10.0&  1.0&200.0&   10$^6$&50\\
32&   1.0& 10.0&200.0&   10$^6$&50\\
33&  10.0&  1.0&200.0&    10$^5$& 1\\
34&  10.0&  1.0&200.0&    10$^5$&50\\
35&   1.0& 10.0&200.0&    10$^5$& 1\\ 
36&   1.0& 10.0&200.0&    10$^5$&50\\
37&   1.0&  1.0& 50.0&     10$^4$& 1\\
38&   1.0&  1.0& 50.0&     10$^4$&50\\
39&   1.0&  1.0& 50.0&    10$^5$&50\\
40&   1.0&  1.0& 50.0&   10$^6$&50\\
41&   1.0&  1.0& 100.0&   10$^6$&2 \\
42&   1.0&  100.0& 100.0&   10$^6$&2 \\
43&   1.0&  10$^3$& 100.0&   10$^6$&2 \\
44&   1.0&  10$^4$& 100.0&   10$^6$&2 \\
\noalign{\smallskip}
\hline
\hline
\end{tabular}\label{Tab5}}
\tablefoot{Column (1) lists the Model number. Columns (2) and (3) list, respectively, the cosmic-ray ionization rates ($\zeta$) and radiation fields ($\chi$) in units of their standard Galactic values ($\zeta_o$ = 5~$\times$~10$^{-17}$ s$^{-1}$ and 1 Draine field~$\equiv$~2.74~$\times$~10$^{-3}$~erg~s$^{-1}$~cm$^{-2}$). Columns (4) and (5) list, respectively the kinetic temperature (T$_{\rm k}$) and hydrogen density (n$_{\rm H}$) of the gas. Column (6) lists the depth of the layer in  A$_V$-units.}\\
\end{table}
 

 For  temperatures $T_{\rm ex}\geq25$~K the derived C$_2$H abundances lie in the range $X$(C$_2$H) $\simeq10^{-7}-10^{-6}$. It is worth noting that these {\rm higher} temperatures are specifically required in order to fit the multiline CO dataset of the {\it outflowing} CND \citep{Vit14}. In particular,   \citet{Vit14} used LTE to conclude that the rotational temperature, $T_{\rm rot}$, of the CO emitting gas in the CND $\simeq40-60$~K. This is also in excellent agreement with the values of the dust temperature required to fit the SED of dust emission in the outflow region out to $r\leq400$~pc:  T$_{\rm dust}\simeq46\pm3$~K \citep{GB14}. 
 Furthermore \citet{Vit14}  used non-equilibrium radiative transfer schemes (RADEX) to conclude that the kinetic temperature of molecular gas traced by CO lines in the outflow is significantly high:  $\geq100$~K.  In contrast, a similar analysis suggests that kinetic temperatures of molecular gas are comparatively a factor $\simeq$2-3 lower in the SB ring.
 
 In order to nail down more accurately how $X$(C$_2$H) changes in the different environments of NGC~1068 
 we would need to independently estimate $T_{\rm ex}$ for C$_2$H by observing at least 1--2 higher $N$ transitions. However, if we assume that the excitation of C$_2$H follows a pattern similar to the rest of molecular species, the observational evidence discussed above suggests that $T_{\rm ex}$ is likely a factor of 2--3 higher in the outflow region compared to the SB ring, and, also, that $\simeq10$~K is a conservative lower limit for $T_{\rm ex}$ in either case.  We will therefore adopt the following reference values for $X$(C$_2$H): $\simeq$ a few 10$^{-8}$  in the SB ring for $T_{\rm ex}\simeq10$~K,  and  $\simeq$ a few 10$^{-7}$ in the outflow region for  $T_{\rm ex}\simeq25-50$~K.

\section{The origin of the C$_2$H abundances}\label{model}

 In this section we aim to account for the range of abundances derived in Sect.~\ref{abun} for C$_2$H in the SB ring and in the outflow region of NGC~1068 using a set of time-dependent gas-grain chemical models, parametrized shock models, and  gas-phase PDR models.

\subsection{Chemical models: description}

 We note that a comprehensive chemical modeling of the C$_2$H molecule is beyond the scope of this observational paper. Moreover,  we are aware of the limitations of having observed only one $N$  transition of the C$_2$H multiplet, leading to an uncertain derivation of the fractional abundance. Hence we can only qualitatively discuss the origin of this emission and its potential role as a tracer of distinct gas components.  We use and augment the
grid of chemical models from \citet{Vit17} run using the UCL\_CHEM time dependent gas-grain chemical model 
\citep{Vit11}, which includes a parametrized shock module \citep{Jim08}, as well as a very small grid of  
PDR models taken from \citet{Bel06} and \citet{Vas10}, run using the UCL\_PDR code. 
We assume for all the models a solar C/O abundance ratio $\sim$ 0.6. Below we briefly present the codes, which are
more extensively described in the above references.

The code UCL\_CHEM \citep[now publicly available at https://uclchem.github.io and published in][]{Hol17} is a time dependent gas-grain chemical model that computes the evolution  of chemical abundances of
the gas and on the ices as a function of time, starting from a diffuse and atomic gas. The code is coupled with the parametrized shock model of  \citet{Jim08}. The code is run in two distinct temporal phases, denoted as  Phase I and Phase II. The initial density in Phase I is $\sim$ 10 cm$^{-3}$. The gas is allowed to collapse and reach a high density by means of a free-fall collapse process.
In this context the collapse is not meant to represent the formation of protostars, but it is simply a way to compute the chemistry of high density gas in a self-consistent way starting from a diffuse atomic gas, i.e., without assuming the chemical composition at its final density.  The temperature during this phase is kept constant at 10 K, and the cosmic-ray ionization rates and radiation fields are at their standard Galactic values ($\zeta_o$ = $5\times10^{-17}$ s$^{-1}$ and 1 Draine field~$\equiv$~2.74~$\times$~10$^{-3}$~erg~s$^{-1}$~cm$^{-2}$, respectively). During Phase I
atoms and molecules are allowed to freeze onto the dust grains and react with each other, forming icy mantles. Non-thermal desorption is also included. In Phase II, for the non-shock models,
UCL\_CHEM computes the chemical evolution of the gas and the icy mantles after either an assumed burst of star formation or an AGN activity episode has occurred. The temperature of the gas increases from 10~K  to a value set by the user, and thermal sublimation from the icy mantles occurs. The chemical evolution of the gas is then followed for 10$^7$ years. In the shock models, Phase II considers a plane-parallel C-Shock propagating with a velocity, $V_{\rm s}$, through the ambient medium.

One of the effects of the AGN activity on the gas is an enhanced cosmic-ray and/or X-ray ionization rate. 
The grid of models we employ from \citet{Vit17} uses the cosmic-ray ionization flux to also simulate an enhancement in 
X-ray flux. Although this approximation has its limitations, in that the X-ray flux ionizes and heats the gas more efficiently
 than cosmic rays, the chemistry arising from these two fluxes should be similar and the chemical trends are the 
same for cosmic rays and X rays for a fixed gas temperature, which is what is assumed in these models \citep[see also][]{Vit14}.

While some of the models from the UCL\_CHEM grid may be considered as representative of PDR models, since 
the radiation field is enhanced enough to penetrate great depths and hence affect the chemistry, we also consider a very small grid of pure gas-phase PDR models at lower gas density. The latter would be more representative of the diffuse and translucent clouds where  C$_2$H has been observed at fractional abundances of $\sim$ 10$^{-8}$  \citep[e.g.][]{Tur99} and where the temperature is highly affected by the UV radiation field. These models were run with a self-consistent PDR model: 
the UCL\_PDR code is a 1 D time dependent PDR code. It was first developed
by \citet{Bel05} and \citet{Bel06} and further augmented by \citet{Bay12} and Priestley et al. (2017, submitted). The code assumes a semi-infinite slab geometry, and computes the chemistry and the temperature as functions of depth and time within the semi-infinite slab.
The chemistry and thermal balance are calculated self-consistently at each depth point into the slab and at each time-step, producing chemical abundances, emission line strengths and gas temperatures as a function of depth and time. 
The code has been benchmarked \citep{Rol07}.

 
 \begin{figure*}
   \centering
    \includegraphics[angle=0,scale=0.85]{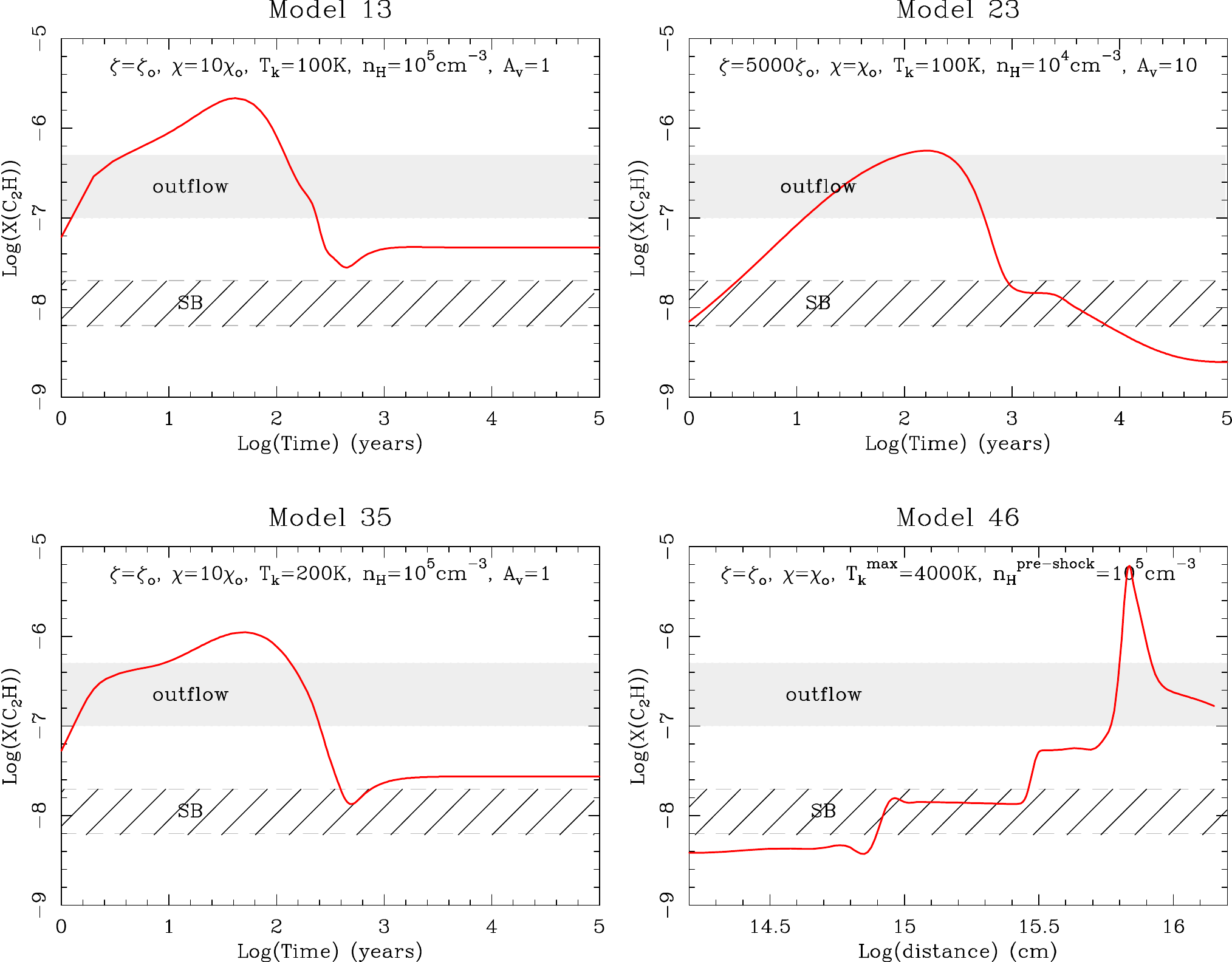}
   \caption{Fractional abundance of C$_2$H for a representative set of chemical models discussed in Sect.~\ref{model} that account for the measured abundances of ethynyl in the SB ring and the outflow region of NGC~1068. The  C$_2$H abundance is represented as a function of time for gas-grain chemical models  (UCL\_CHEM)  number 13, 23 and 35, and as a function of distance for shock model number 46. This scale corresponds to the distance that the shock travels up to the dissipation length, which is the length scale over which the shock dissipates, beyond which the model is no longer applicable \citep[see][]{Jim08}. The main parameters of the models, extensively described in Sect.~\ref{model} and listed in Tables~\ref{Tab4}, \ref{Tab5} and \ref{Tab6}, are indicated in each panel. We highlight the range of C$_2$H abundances measured in the SB ring ($\simeq$ a few 10$^{-8}$ for $T_{\rm ex}\simeq10$~K) and in the outflow region ($\simeq$ a few 10$^{-7}$ for  $T_{\rm ex}\simeq25-50$~K). }
   \label{Fig7}
    \end{figure*}
    

\begin{table}
\caption{Grid of parametrized shock models.}
\centering
\resizebox{0.4\textwidth}{!}{
\begin{tabular}{cccccc}
\hline
\hline
\noalign{\smallskip}
Model & $\zeta$ & $\chi$ & T$_{\rm max}$ & n$_{\rm H_{pre-shock}}$  & \\
\noalign{\smallskip}
 & ($\zeta_0) $ & (Draine field) & (K) & (cm$^{-3}$) & \\
 \noalign{\smallskip}
 \hline
 \noalign{\smallskip}
45&   1.0&  1.0& 2200 & 10$^4$ & \\ 
46&   1.0&  1.0& 4000 & 10$^5$ & \\  
47&   1.0&  1.0& 4000 & 10$^6$ & \\ 
\noalign{\smallskip}
\hline
\hline
\end{tabular}\label{Tab6}}
\tablefoot{Column (1) lists the Model number. Columns (2) and (3) list, respectively, the cosmic-ray ionization rates ($\zeta$) and radiation fields ($\chi$) in units of their standard Galactic values ($\zeta_o$ = 5~$\times$~10$^{-17}$ s$^{-1}$ and 1 Draine field~$\equiv$~2.74~$\times$~10$^{-3}$~erg~s$^{-1}$~cm$^{-2}$). Columns (4) and (5) list, respectively the maximum kinetic temperature (T$_{\rm max}$) and pre-shock hydrogen density (n$_{\rm H_{pre-shock}}$) of the gas.}\\
\end{table}
 

\subsection{Chemical models: results}

Tables~\ref{Tab4}, \ref{Tab5} and \ref{Tab6} list, respectively,  the parameters of the grid of UCL\_PDR,  UCL\_CHEM and shock models we considered. All our models have been computed with solar abundances.  

We note that none of the pure gas-phase low density (10$^3$ cm$^{-3}$) PDR models (Models 48-51; Table~\ref{Tab4}), at any time or depth into the cloud, succeed in achieving a fractional abundance of C$_2$H larger 
than 10$^{-9}$. This is consistent with the results of previous PDR models.  For example,  \citet{Tur99} 
succeeded in fitting the C$_2$H abundance observed in some translucent clouds (10$^{-8}$) with steady state 
models but only by assuming a high non-solar C/O ratio ($\sim$ 0.9). In denser PDRs \citet{Cua15} found that a C$_2$H abundance of 10$^{-8}$ can in fact only be matched by a model with a gas density of $>$ 10$^6$ cm$^{-3}$, a radiation field of $>$ 10$^4$ Draine field and only at a depth $<$ 2 mag. In other words, only an extremely 
dense but thin layer of gas can sustain steadily such a high abundance of C$_2$H.     
Such scenario is depicted by some of our models from the UCL\_CHEM grid and they will be discussed later. 

We now discuss the output of Models 1--44 (Table~\ref{Tab5}) in the light of the abundances derived for the 
outflow region. If we assume the temperature of the gas is $>$ 50 K then 
none of the UCL\_CHEM models at equilibrium succeeds in producing enough C$_2$H, with maximum 
abundances that reach at most $\simeq$~3$\times$~10$^{-8}$. However, there is a class of models, namely 
Models 13, 23, 35, where higher abundances, comparable to (or higher 
than) those observed, can be reached at {\it very early times}. These models are all characterized by either a 
higher (by a factor of 10) than standard UV radiation field, or by a very high  cosmic-ray ionization 
rate (a factor of 5000 higher than standard); the latter would account for the influence of enhanced X-ray irradiation. All models have at least a density of 10$^4$ cm$^{-3}$. 
We plot in Fig.~\ref{Fig7} the temporal evolution of C$_2$H for these models. The increase and decrease of C$_2$H are due, respectively, to 
photodissociation of C$_2$H$_2$ and ion-neutral reaction of C$_2$H with C$^+$.  
While we can not differentiate between a UV or a cosmic-ray/X-ray enhanced field, all these models are consistent with those of molecular interfaces between outflows and ambient clouds in high-mass star-forming regions in our 
Galaxy \citep{Cod06, Kem10, San16}, where visual extinctions are low ($\leq$ 2 mag), gas densities, temperatures and radiation fields are high ($>$ 10$^4$ cm$^{-3}$,  $>$ 100 K, and  $>$ 10 Draine field, respectively), and the gas is short-lived ($\simeq$ 10$^{2-3}$~yr). We interpret that 
these transient conditions in the UV/X-ray irradiated gas may be due to non-dissociative shocks associated with the highly turbulent interface between the outflow and the 
molecular clouds. While it is true that Model 23 has a high visual extinction, due to the very high cosmic-ray ionization 
rate the chemistry of the gas is effectively the same as that of an interface. 
Note that although timescales may be very short, the continual erosion of dense material by the wind or outflow would resupply the 
interface, so a nearly steady state may persist in the UV/X-ray illuminated gas across the shock front.

Models 1--44, although partly mimicking the effects of non dissociative shocks, did not model properly the passage of the shock. In order to test our hypothesis 
that the C$_2$H abundances can be reproduced by the erosion of icy mantles due to the passage of non dissociative shocks, we also 
include in our grid three shock models (45--47; Table~\ref{Tab6}). In these models we assumed  an average shock velocity  $V_{\rm s}\simeq40$~km~s$^{-1}$. The pre-shock density and the shock velocity determine the maximum temperature reached (see Table~\ref{Tab6}).
We plot the results of Model 46 in Fig.~\ref{Fig7}. 
It is clear that the C$_2$H abundance does indeed rise as the shock penetrates the dissipation length of the gas, and it is close to that found by 
our LTE analysis if one assumes a 100 K gas, which, especially in the outflow region, is a reasonable gas temperature \citep{GB14, Vit14}.

With all the necessary caveats in mind, these results suggest that the inferred high abundances of C$_2$H in the outflow could represent the first 
evidence of a {\it chemical} fingerprint of an outflow interface region in an external galaxy. 
The scenario described above, where UV/X-ray radiation and shocks seem to act in concert to shape the chemistry of the outflow interface region in 
NGC~1068, is also similar to the one supporting the existence of mechanical dominated regions (MDRs), a variant of PDRs where mechanical 
action is invoked to explain the spectral energy distribution of CO in some star-forming and active galaxies 
\citep{Mei13, Ros14, Ros15}.

We now examine the results of Models 1--44 in the light of the abundances derived for the SB ring.  While the derived fractional abundances as a 
function of temperatures are essentially the same as in the outflow spot, we note that in the SB ring the derived temperatures are comparatively 
lower (Sect.~\ref{abun}) and, since C$_2$H may be sub-thermally excited, the derived fractional abundances may nevertheless correspond to 
those derived for a gas at a higher (than 10 K) excitation temperature. In this case, the abundances that our models need to match are more of 
the `canonical' order of $\sim$ 10$^{-8}$. We first look at the models run with a temperature of 50 K (Models 37--40). In fact Model 37, which 
represents a `canonical'  PDR, with Galactic UV and cosmic-ray ionization fields, a typical molecular cloud gas density of 10$^4$ cm$^{-3}$, a 
visual extinction of 1 mag and a temperature of 50 K, provides us with a match with an abundance of  10$^{-8}$. This may therefore indicate that 
C$_2$H is simply tracing the gas most exposed to star formation in a region where we know star formation is very active. Other models match 
this abundance at equilibrium as well: in particular Models 1, 11, 13, and 16, {\it all} at A$_V$ = 1 mag (hence PDR-like), and differing by a factor 
of 10 in either radiation field, cosmic-ray ionization field and gas density from Model 37, hence all `reasonable' PDR models depicting gas 
affected by active star formation, as expected in a SB ring. However we note that all these models are run at gas temperatures of 100 K. 
Moreover, of Models 41--44, mimicking the best fit models from \citet{Cua15}, also Models 42--44 can reproduce the observed abundance of 1-3$\times$10$^{-8}$, making it difficult to determine whether C$_2$H is tracing a very dense gas subjected to an enhanced radiation field, or a 
more canonical molecular cloud gas. Regardless, the C$_2$H must come from a very low visual extinction gas which means that the SB ring 
must be at least partly dominated by PDR gas. 
Hence for the SB ring, C$_2$H may simply be tracing star formation activity. 

Further confirmation of all the favored scenarios discussed above would need a more extensive chemical modeling that  explores, for example, a 
wider range of initial conditions, including initial elemental abundances differing from solar.

\section{Summary and conclusions}

We have used ALMA to map the emission of the hyperfine multiplet of C$_2$H($1-0$) and its underlying continuum emission at 86.3~GHz in the 
central $r\simeq35\arcsec$(2.5~kpc)-region of the disk of the nearby ($D=14$~Mpc) Seyfert 2 galaxy NGC~1068  with a spatial resolution $1\farcs0\times0\farcs7$~($\simeq50-70$~pc).  In an attempt to explain the fractional abundances derived for C$_2$H in the different spatially-
resolved environments of the galaxy we have developed a set of time-dependent chemical models.

We summarize the main results of our study as follows:

\begin{itemize}

\item

We detected widespread C$_2$H emission in the disk of NGC~1068. Most ($\simeq65\%$) of the spatially-integrated C$_2$H line flux stems 
from the $r\simeq1.3$~kpc SB molecular ring. The SB ring concentrates the bulk of the recent massive star formation in the disk of the galaxy as 
traced by the strong Pa$\alpha$ emission of  H{\small II} regions imaged by HST. The brightest C$_2$H emission originates nevertheless from 
the $r\simeq200$~pc {\it  outflowing} CND. We also detected significant C$_2$H emission bridging the CND with the outer disk in a region that 
probes the interface between the molecular disk and the ionized gas outflow out to $r\simeq400$~pc.

\item

 The line ratios measured between the hyperfine components of the C$_2$H multiplet detected throughout the disk of NGC~1068 indicate that their emission is mostly optically thin. We derived the column densities of C$_2$H   assuming  LTE conditions and a set of excitation temperatures  constrained by the previous multiline CO studies of the galaxy.   
 
 \item
 
 Maps of the dust continuum emission obtained at 349~GHz by ALMA were used to calculate the H$_2$ gas column densities. These were used in combination  with the C$_2$H column densities derived from LTE to obtain the fractional abundances of this species in different regions. Our estimates range from $X$(C$_2$H)~$\simeq$~a few 10$^{-8}$  in the SB ring up to $X$(C$_2$H)~$\simeq$~a few 10$^{-7}$ in the outflow.

\item 
We performed a preliminary qualitative chemical analysis to determine the 
origin of the high abundance of C$_2$H in the outflow and in the SB ring. 
In both regions we find that the gas is dense ($\geq$ 10$^4$ cm$^{-3}$), confirming previous results \citep{GB14, Vit14}.
\item

In the outflow we find that fractional abundances of $X$(C$_2$H)~$\simeq$ a few 10$^{-7}$ can only be reached at {\it very early} times in models where dense molecular gas is heavily irradiated by UV or X-ray photons. In particular, these models are consistent with those of molecular interfaces between outflows and ambient clouds, where visual extinctions are low ($\leq$ 2 mag), gas densities, temperatures and UV fields or cosmic-ray/X-ray fields are high ($>$10$^4$ cm$^{-3}$,  $>$~100 K, and $>$~10 Draine field, or$>$~5000~$\zeta_o$, respectively),  and the gas is short-lived ($\simeq$ 10$^{2-3}$yr). 

\item 
We interpret that transient conditions in the outflow may be due to  UV/X-ray irradiated non-dissociative shocks associated with the highly turbulent interface between the outflow and the molecular clouds. Although timescales may be very short, the continual erosion of dense material by the wind or jet resupplies the interface, so a nearly steady state persists \citep[e.g.][]{Cod06}. Shock models seem to confirm that C$_2$H can also be enhanced due to icy mantle sputtering.

\item 
In the SB ring on the other hand an abundance of $X$(C$_2$H)~$\simeq$~a few~10$^{-8}$ is either tracing  molecular gas typical of the `skin' of Galactic giant molecular clouds, or a much denser `skin' subjected to an enhanced UV field. Both scenarios are consistent with active star formation.

\end{itemize}

Our chemical modeling is not extensive and has not covered a large enough parameter space of initial conditions to exclude the validity of alternative scenarios. In particular,  we cannot exclude that the hydrocarbons chemistry is incomplete in our models or that PAH chemistry (not included) has a significant effect on the gas-phase chemistry of C$_2$H \citep[e.g][]{Pet05}; equally a small difference in the C/O ratio in the adopted  initial abundances of our models could change our conclusions as it would affect the abundance of all hydrocarbons \citep{Cua15}. Once further transitions of C$_2$H are obtained, a more comprehensive modeling would be needed to shed light on the origin of the significantly high abundances of C$_2$H in NGC~1068.

\begin{acknowledgements}
         We acknowledge the staff of ALMA in Chile and the ARC-people at IRAM-Grenoble in France 
for their invaluable help during the data reduction process. This paper
makes use of the following ALMA data: ADS/JAO.ALMA$\#$2013.1.00055.S and $\#$2011.0.00083.S.
ALMA is a partnership of ESO (representing its member states), NSF (USA)
and NINS (Japan), together with NRC (Canada) and NSC and ASIAA (Taiwan),
in cooperation with the Republic of Chile. The Joint ALMA Observatory is
operated by ESO, AUI/NRAO and NAOJ. The National Radio Astronomy
Observatory is a facility of the National Science Foundation operated under cooperative
agreement by Associated Universities, Inc. 
We used observations made
with the NASA/ESA Hubble Space Telescope, and obtained from the Hubble
Legacy Archive, which is a collaboration between the Space Telescope Science
Institute (STScI/NASA), the Space Telescope European Coordinating Facility
(ST-ECF/ESA), and the Canadian Astronomy Data Centre (CADC/NRC/CSA).     
 SGB and AU acknowledge support from the Spanish MINECO and FEDER funding grants AYA2016-76682-C3-2-P, AYA2016-79006-P and 
 ESP2015-68964-P. CRA acknowledges the Ram\'on y Cajal Program of the Spanish Ministry of Economy and Competitiveness through project 
RYC-2014-15779 and the Spanish Plan Nacional de Astronom\'{\i}a y Astrof\'{\i}sica under grant AYA2016-76682-C3-2-P. AAH 
acknowledges support from the Spanish MINECO and FEDER funding grant AYA2015-6346-C2-1-P.

\end{acknowledgements}

\end{document}